\documentclass[english,aps,pra,superscriptaddress,reprint,twocolumns]{revtex4-1}
\usepackage[english]{babel}
\usepackage[T1]{fontenc}
\usepackage{amsmath,amssymb,graphicx,mathrsfs,xcolor,babel,textcomp,esint,ulem}
\usepackage[unicode=true,pdfusetitle,bookmarks=true,bookmarksnumbered=false,bookmarksopen=false,breaklinks=true,pdfborder={0 0 0},backref=false,colorlinks=true]{hyperref}

\begin{document}

\title{Applicability of coupling strength estimation for linear chains of
restricted access}

\author{He Feng}
\affiliation{Shanghai Advanced Research Institute, Chinese Academy of Sciences, Shanghai 201210, China}
\affiliation{University of Chinese Academy of Sciences, Beijing 100049, China}
\author{Tian-Min Yan}
\email{yantm@sari.ac.cn}
\affiliation{Shanghai Advanced Research Institute, Chinese Academy of Sciences, Shanghai 201210, China}
\author{Y. H. Jiang}
\email{jiangyh@sari.ac.cn}
\affiliation{Shanghai Advanced Research Institute, Chinese Academy of Sciences, Shanghai 201210, China}
\affiliation{University of Chinese Academy of Sciences, Beijing 100049, China}
\affiliation{ShanghaiTech University, Shanghai 201210, China}

\begin{abstract}
  The characterization of an unknown quantum system requires the Hamiltonian
  identification. The full access to the system, however, is usually
  restricted, hindering the direct retrieval of relevant parameters, and a
  reliable indirect estimation is usually required. In this work, the
  algorithm proposed by Burgarth et al. [Phys. Rev. A {\bfseries{79}}, 020305
  (2009)], which allows estimating the coupling strengths in a linear chain by
  addressing only one end site, is further investigated. The scheme is
  numerically studied for states with chain structure, exploring its
  applicability against observational errors including the limited
  signal-noise ratio and the finite spectral width. The spectral distribution
  of the end state is shown to determine the applicability of the method, and
  reducing the loss from truncated spectral components is critical to
  realizing the robust reconstruction of coupling strengths.
\end{abstract}
\maketitle

\section{Introduction}

The accurate control of the Hamiltonian of designed systems is always a
prerequisite to carry out desired tasks, e.g., quantum computation
{\cite{ladd_quantum_2010,bassett_quantum_2012}} quantum communication
{\cite{bose_quantum_2003}} and quantum metrology
{\cite{giovannetti_quantum_2006,giovannetti_advances_2011}}. The systems are
usually microscopic structures which are delicately engineered to realize
specific functions. To verify and benchmark these fabricated structures, the
characterization of the system via Hamiltonian identification is desired
{\cite{cole_hamiltonian_2015}}. Usually, the dynamics within the system are
extremely complicated, and the probing access is often restricted. A
delicately devised identification scheme is supposed to allow the sensible
estimation of unknown parameters, reconstructing the Hamiltonian indirectly
based on partially available information, e.g., the {\itshape{a priori}}
knowledge about the structure of the quantum network, the initial state, or an
accessible subset of observables, {\itshape{et al.}}.

Under the challenge of complicated dynamics and restricted addressing
resources, various identification algorithms have been developed aiming at
experimental realizations. In a many-body system, the dynamical decoupling
technique, which simplifies the problem by decoupling each pair of qubits from
the rest, allows for the Hamiltonian identification with arbitrary long-range
couplings between qubits {\cite{wang_hamiltonian_2015}}. Besides, the dynamics
can be altered by tuning the control pulse that is applied to the probe spin,
improving the precision of estimation scheme
{\cite{kiukas_remote_2017,liu_quantum_2017}}. In a network of limited access,
the identification scheme is intended to bridge the Hamiltonian parameters and
the observables from accessible subsystem that are relatively easy to measure.
The system realization theory is proposed for temporal record of the
observables of a local subsystem
{\cite{zhang_quantum_2014,zhang_identification_2015}}, which has been
experimentally demonstrated on a liquid nuclear magnetic resonance quantum
information processor {\cite{hou_experimental_2017}}. Zeeman marker protocol
shows that local field-induced spectral shifts can be used to estimate
parameters in spin chains or networks {\cite{burgarth_evolution-free_2017}}.
Also, when the graph infection rule is satisfied, the similar parameter
estimation is also available utilizing the spectral information retrieved from
a partially accessible spin
{\cite{burgarth_coupling_2009,burgarth_indirect_2009}}.

In this work, we re-examine the estimation algorithm proposed in
{\cite{burgarth_coupling_2009}}, where the reconstruction of coupling
strengths in a linear chain without the full access is considered. The
procedure, probing the global properties from a local site, is similar to the
estimation of spring constants in classical-harmonic oscillator chains. By
accessing only the end of the linear chain, the algorithm allows deducing all
coupling strengths from the data of associated spectral information.
Nevertheless, the errors of the input data are inevitable and may
significantly influence the reconstruction results. In this work, we focus on
the applicability of the algorithm when the acquired initial data deviate from
the actual values. The simulations show that the spectral distribution is an
important indicator of the applicability of the algorithm. Examining the
spectral distribution, the increasing number of spectral components that
approach zero are likely to fail the reconstruction. The applicability depends
on both the properties of individual system and conditions of measurement. The
work is organized as followings. In Sec. \ref{sec:theory}, the recursive
relations among spectral coefficients and coupling strengths from adjacent
sites are derived. In Sec. \ref{sec:applicability}, the robustness and the
availability of the algorithm are discussed under different conditions when
input errors are introduced.

\section{Theory of coupling strength estimation}\label{sec:theory}

\begin{figure}[h]
  \includegraphics[scale=1]{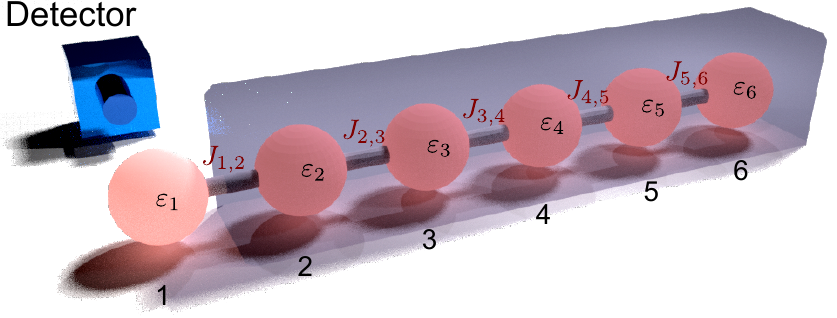}
  \caption{Schematics of parameter estimation for a chain of $N$ states.
  Assuming that the access of the chain is restricted, only state $|1
  \rangle$, the left-most site of the chain, can be measured by the detector
  and all the rest (states $|2 \rangle$, $|3 \rangle$, ..., $|N \rangle$) is
  concealed by a blackbox. Using the scheme of parameter estimation, however,
  all coupling strengths, $J_{i, i + 1}$, can be estimated if the spectral
  information of $|1 \rangle$ is available. \label{fig:chain-scheme}}
\end{figure}

We consider the state transfer within a system governed by the generic
Schr{\"o}dinger equation,
\begin{equation}
  \dot{c}_i = - \mathrm{i} \varepsilon_i c_i - \mathrm{i} \sum_{k \sim i}
  J_{i, k} c_k, \label{eq:schroedinger-eq}
\end{equation}
with $c_i$ the amplitude of state $|i \rangle$, $\varepsilon_i$ the energy and
$J_{i, k}$ the coupling strength. The symbol "$\sim$" means the two states
are coupled. With amplitude $c_i = \langle i| \Psi (t) \rangle$ represented
spectrally, $c_i = \sum_n C_{i, n} \mathrm{e}^{- \mathrm{i} \lambda_n t}$,
where $C_{i, n} = \langle  i| \lambda_n \rangle \langle \lambda_n | \psi (0)
\rangle$ is the spectral coefficient of mode $n$. The relation of $C_{i, n}$
among coupled states reads
\begin{equation}
  C_{i, n} = \frac{1}{\lambda_n - \varepsilon_i}  \sum_{k \sim i} J_{i, k}
  C_{k, n} . \label{eq:relation-of-spec-coef-from-dc}
\end{equation}
The relation allows estimating the parameters using the information retrieved
from a subset of states under the restricted-access condition. The simplest
example is a linear chain as shown in Fig. \ref{fig:chain-scheme}, where only
the leftmost state $|1 \rangle$ is accessible. Given an $N$-state chain with
energies $\varepsilon_i$ known, all coupling strengths $J_{i, i + 1}$ can be
reconstructed using the spectral information simply read from state $|1
\rangle$.

The feasibility of the scheme can be shown by the repetitive use of Eq.
(\ref{eq:relation-of-spec-coef-from-dc}) and the normalization condition
$\langle i|i \rangle = 1$. Given the eigenmode $n$, the $C_{1, n}$ of
left-most state $|1 \rangle$ allows deriving $C_{2, n}$ of the next one,
\begin{equation}
  C_{2, n} = \frac{\lambda_n - \varepsilon_1}{J_{12}} C_{1, n} .
  \label{eq:C-recursion-1}
\end{equation}
Given that only $J_{i, k}$ with $k = i \pm 1$ are nonvanishing in a linear
chain, the repetitive use of Eq. (\ref{eq:relation-of-spec-coef-from-dc})
yields coefficients of subsequent states recursively,
\begin{equation}
  C_{i + 1, n} = \frac{(\lambda - \varepsilon_i) C_{i, n} - J_{i - 1, i} C_{i
  - 1, n}}{J_{i, i + 1}} . \label{eq:C-recursion-2}
\end{equation}
The denominator $J_{i, i + 1}$ should be non-zero, since a "broken" coupling
forbids the retrieval of the information on the further side. Eqs.
(\ref{eq:C-recursion-1}) and (\ref{eq:C-recursion-2}) allow for the recursive
evaluation of $C_{i + 1, n}$ from $C_{i, n}$ and $C_{i - 1, n}$.

On the other hand, the normalization condition $\langle i|i \rangle = 1$
should be satisfied. With $C_{i, n} = \langle  i| \lambda_n \rangle \langle
\lambda_n | \psi (0) \rangle$, we have $\langle i| i \rangle = \sum_n \langle 
i| \lambda_n \rangle \langle  \lambda_n |i \rangle = \sum_n | \frac{C_{i,
n}}{\langle \lambda_n | \psi (0) \rangle} |^2 = 1.$ The unknown denominator
$\langle \lambda_n | \psi (0) \rangle$ depends on the concrete form of initial
state $| \psi (0) \rangle$. If the system is initially prepared by populating
state $|1 \rangle$ only, as considered in our case, $| \psi (0) \rangle = |1
\rangle$, the denominator reads $| \langle \lambda_n | \psi (0) \rangle |^2 =
| \langle \lambda_n |1 \rangle |^2 = C_{1, n}$ and the normalization condition
becomes
\begin{equation}
  \langle i|i \rangle = \sum_n \frac{| C_{i, n} |^2}{C_{1, n}} = 1.
  \label{eq:normalization}
\end{equation}
With only state $|1 \rangle$ accessible, the information of $|1 \rangle$, $c_1
(t) = \sum_n C_{1, n} \mathrm{e}^{- \mathrm{i} \lambda_n t}$, is supposed to
be acquired by measurement, providing the input values of $C_{1, n}$ and
$\lambda_n$ for further parameter estimation. The measured $C_{1, n}$ should
be normalized by Eq. (\ref{eq:normalization}). Next, substituting Eq.
(\ref{eq:C-recursion-1}) into Eq. (\ref{eq:normalization}) with $i = 2$, the
normalization condition $\langle 2|2 \rangle = 1$ yields
\begin{equation}
  J_{1, 2} = \sqrt{\sum_n^N | \lambda_n - \varepsilon_1 |^2 C_{1, n}}
  \label{eq:J-recursion-1}
\end{equation}
and the normalized $C_{2, n}$ using Eq. (\ref{eq:C-recursion-1}). With the
evaluated $J_{i - 1, i}$, $C_{i, n}$ and $C_{i - 1, n}$, Eq.
(\ref{eq:normalization}) generates the $J_{i, i + 1}$ for $i \geqslant 2$,
\begin{equation}
  J_{i, i + 1} = \sqrt{\sum_n^N \frac{| (\lambda_n - \varepsilon_i) C_{i, n} -
  J_{i - 1, i} C_{i - 1, n} |^2}{C_{1, n}}} . \label{eq:J-recursion-2}
\end{equation}
Hence, all coupling strengths can be recursively evaluated based on the above
derived Eqs . (\ref{eq:C-recursion-1}), (\ref{eq:C-recursion-2}),
(\ref{eq:J-recursion-1}) and (\ref{eq:J-recursion-2}).

In a realistic experimental setup, the above scheme works for any system that
can be reduced to the form equivalent to Eq. (\ref{eq:schroedinger-eq}). In
{\cite{burgarth_indirect_2009}}, the method is applied to the spin chain
described by Heisenberg Hamiltonian,
\begin{equation}
  \hat{H} = \sum_i^{N - 1} J_{i, i + 1} (\sigma_i^+ \sigma_{i + 1}^- +
  \sigma_i^- \sigma_{i + 1}^+ + \Delta \sigma_i^z \sigma_{i + 1}^z)
  \label{eq:heisenberg-hamiltonian}
\end{equation}
with $\Delta$ the anisotropy. Assuming the system is prepared within the
single excitation section, the subsequent evolution under the Hamiltonian
(\ref{eq:heisenberg-hamiltonian}) is still restricted to the single excitation
sector. The time evolution is mapped to the generic Schr{\"o}dinger equation
for a single particle, Eq. (\ref{eq:schroedinger-eq}), where the energies are
given by $\varepsilon_i = \Delta \left[ \sum_{j = 1}^{N - 1} J_{j, j + 1} - 2
(J_{i, i + 1} + J_{i - 1, i}) \right]$. Therefore, the above mentioned states
$\{ |i \rangle \}$, in this case, have been mapped to spatial sites. The
values of $C_{1, n}$ are embedded in the reduced density matrix of $|1
\rangle$, which can be experimentally obtained by quantum state tomography
{\cite{burgarth_indirect_2009}}. Here, we are not concerned with the specific
realization of the measurement, Hence, the input variables $\lambda_n$ and
$C_{1, n}$ are assumed to be readily reachable, but they are not necessarily
guaranteed to be completely precise, as will be discussed later.

\begin{figure}[h]
  \includegraphics[scale=1]{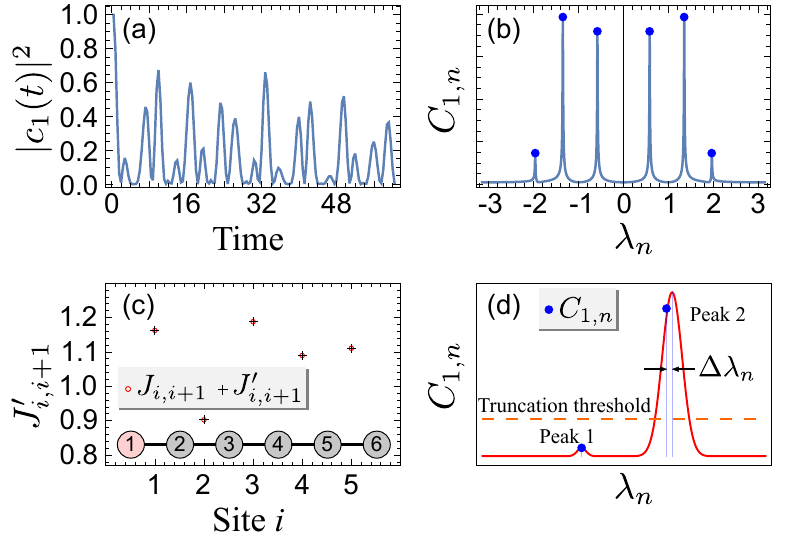}
  \caption{Reconstruction of $J_{i, i + 1}$ for a chain of $N = 6$ and
  $\varepsilon_i = 0$. Panel (a) shows the temporal evolution of population on
  site 1, and (b) shows its spectral distribution that provides $\lambda_n$
  and $C_{1, n}$ as required by the reconstruction algorithm. In (c), applying
  the parameter estimation, the five unknown coupling strengths are estimated.
  The estimated values $J_{i, i + 1}'$ show good agreement with the actual
  values $J_{i, i + 1}$. Panel (d) presents the possible errors with the input
  values $\lambda_n$ and $C_{1, n}$ that may hamper the reconstruction
  procedure. For peak 1, the $C_{1, n}$ below the threshold during the
  measurement is simply truncated. For peak 2, the broadening of the spectral
  peak results in the deviation of the measured eigenvalue $\lambda_n'$ from
  the actual $\lambda_n$.\label{fig:demo-reconstruction}}
\end{figure}

As a simple example, the estimation of $J_{i, i + 1}$ in a chain with six
sites is demonstrated in Fig. \ref{fig:demo-reconstruction}. Since the
influence from disorder is not of concern in this work, all energies
$\varepsilon_i$ are zero, as will also be considered in the following
discussions. With only site $|1 \rangle$ accessible, the time evolution of
state $|1 \rangle$ [Fig. \ref{fig:demo-reconstruction}(a)] is supposed to be
detectable. In the associated spectral distribution [Fig.
\ref{fig:demo-reconstruction}(b)], the position and the height of each
spectral peak provide $\lambda_n$ and $C_{1, n}$, respectively, as required as
input values by the estimation scheme. Performing the evaluation using Eqs .
(\ref{eq:C-recursion-1}), (\ref{eq:C-recursion-2}), (\ref{eq:J-recursion-1})
and (\ref{eq:J-recursion-2}) recursively, we successfully reconstruct all the
five unknown coupling strengths $J_{i, i + 1}$, as shown in Fig.
\ref{fig:demo-reconstruction}(c).

\section{Influence from errors of input variables}\label{sec:applicability}

During the actual measurement, the finite instrumental resolution, the limited
signal-noise ratio and other perturbances may hinder the precise acquisition
of initial input $\lambda_n$ and $C_{1, n}$. Therefore, the robustness of the
algorithm against the deviations of these input values is critical to the
success of the parameter estimation. The influence from imprecision of
measured $C_{1, n}$ has been mentioned in {\cite{burgarth_coupling_2009}},
showing rather robust performance against small deviations. When the finite
signal-noise ratio of $C_{1, n}$ detection is considered, the even worse
situation occurs when partial information are lost due to the truncation of
small values. As is shown in Fig. \ref{fig:demo-reconstruction}(d), the finite
signal-noise ratio sets the truncation threshold. As the value of $C_{1, n}$
for peak 1 is below the line representing the threshold, the corresponding
$C_{1, n}$ is missing. Besides the errors in $C_{1, n}$ which are encoded in
heights of spectral peaks, the $\lambda_n$ that are read from positions of
peaks are also susceptible to errors during the measurement. As is shown by
peak 2 in (d), the nonvanishing spectral width caused by either the finite
instrumental resolution or the fluctuation during the measurement may
contribute to the uncertainty $\Delta \lambda_n$. The imprecise $\lambda_n$
also deteriorate the performance of \ $J_{i, i + 1}$-reconstruction.

We take the example of the simplest parametric setup, a chain of 100 sites
with identical coupling strengths $J_{i, i + 1} = 1$ and energies
$\varepsilon_i = 0$, \ to show how the input values influence the results.
Under the ideal condition without any deviations of $\lambda_n$ and $C_{1,
n}$, the simulated results show the $J_{i, i + 1}$ can be correctly recovered
for a long chain (tested up to thousands sites). In a realistic experiment,
however, the spectral peaks of small-valued $C_{1, n}$ may not be well
resolved, or even simply be truncated. For the $n$th mode, when $C_{1, n}$ is
zero, from Eqs. (\ref{eq:C-recursion-1}) and (\ref{eq:C-recursion-2}), all
$C_{i, n}$ vanish, leading to the zero component in Eq.
(\ref{eq:J-recursion-2}) without the associated contribution; and also, Eq.
(\ref{eq:J-recursion-2}) is not bothered by the numerical difficulty of
division by zero. However, with the less number of observed peaks than the
actual number, the lost information does influence the reconstructed results.

\begin{figure}[h]
  \includegraphics[scale=1]{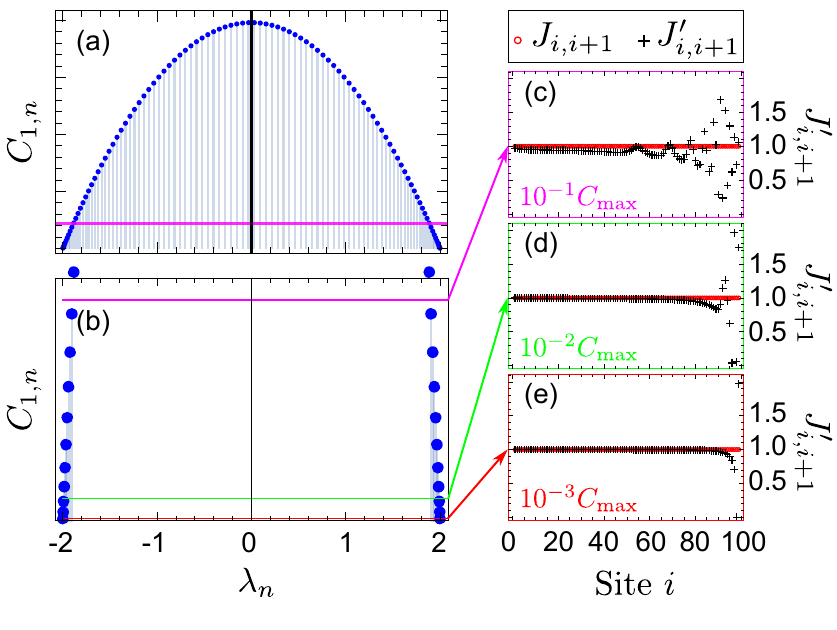}
  \caption{For a homogeneous linear chain of $N = 100$, panel (a) shows the
  spectral distribution $C_{1, n}$ versus $\lambda_n$. The region below $10^{-
  1} C_{\text{max}}$, as indicated by purple line in (a), is zoomed as shown
  in panel (b). Assuming the $C_{1, n}$ can be resolved up to $10^{- 1}$
  (purple), $10^{- 2}$ (green) and $10^{- 3}$ (red) of $C_{\text{max}}$, all
  the input data $(\lambda_n, C_{1, n})$ with $C_{1, n}$ below the threshold
  are truncated. The corresponding estimated coupling strengths $J_{i, i +
  1}'$ are shown in (c), (d) and (e), respectively, comparing with the
  original coupling strengths $J_{i, i + 1}$.\label{fig:C1n-influence}}
\end{figure}

The influence from the truncation is shown in Fig. \ref{fig:C1n-influence}.
For a homogeneous linear chain with all $\varepsilon_i = \varepsilon$ and
$J_{i, i + 1} = J$, the eigenvalues are given by $\lambda_n = \varepsilon + 2
J \cos [\pi n / (N + 1)]$ for $n = 1, \ldots, N$. The element of associated
eigenvector for the $i$th site read $C_{i, n} = \sin [n \pi i / (N + 1)]$. For
the probing state $|1 \rangle$, $C_{1, n} = \sin [n \pi / (N + 1)]$, as is
shown by the spectral distribution in Fig. \ref{fig:C1n-influence}(a). The
$C_{1, n}$ of small values are around both the far ends along the
$\lambda_n$-axis, where the data below a given threshold are truncated if a
finite signal-noise ratio is specified. For different truncation thresholds as
indicated in Fig. \ref{fig:C1n-influence}(b), the estimated results of $J_{i,
i + 1}$ are compared in Fig. \ref{fig:C1n-influence}(c)-(e).

Assuming that the maximum value of $C_{1, n}$ is $C_{\text{max}}$, when the
threshold is $10^{- 1} C_{\text{max}}$ with ten pairs of $C_{1, n}$ truncated,
significant deviation appears from $i = 50$. When the data below $10^{- 2}
C_{\text{max}}$ are truncated with three pairs of $C_{1, n}$ missing, the
$J_{i, i + 1}$ can be correctly reconstructed up to $i = 80$. Further, when
only one pair of data are truncated below the threshold $10^{- 3}
C_{\text{max}}$, the deviation only appear at the rightmost sites of the
chain. The results suggest the complete data of $C_{1, n}$ should be important
to the success of the algorithm. More intriguingly, though all $C_{1, n}$
contribute to the evaluation of each $J_{i, i + 1}$, when some components are
missing, the chain can still be recovered to some extent instead of failing
the reconstruction as a whole, which shows the robustness of the algorithm.

\begin{figure}[h]
  \includegraphics[scale=1]{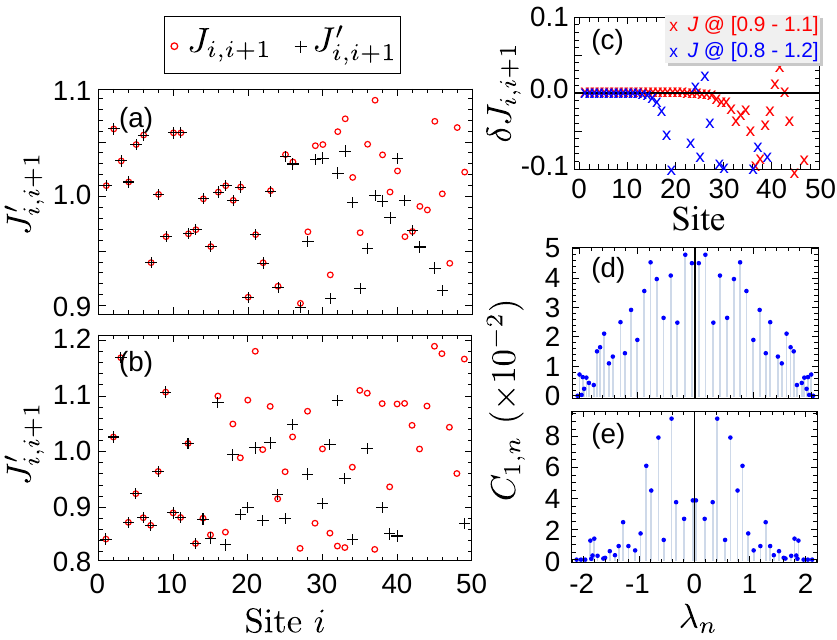}
  \caption{Estimation of $J_{i, i + 1}$ in a chain of $N = 50$ when $C_{1, n}$
  is resolved up to $10^{- 3}$. The $J_{i, i + 1}$ are randomly distributed in
  range (a) $[0.9, 1.1]$ \ and (b) $[0.8, 1.2]$. The errors $\delta J_{i, i +
  1} = J'_{i, i + 1} - J_{i, i + 1}$ for the two different intervals are
  compared in panel (c). The different parametric ranges lead to distinguished
  spectral distributions, as shown in (d) and
  (e).\label{fig:J-range-influence}}
\end{figure}

Next, we consider the estimation when $J_{i, i + 1}$ vary with $i$, as we
desire in practice. Without loss of generality, the $J_{i, i + 1}$ are
randomly chosen within a given interval, i.e., the disorder induced by $J_{i,
i + 1}$. It is shown that the interval is highly relevant to the performance
of the estimation. Figs. \ref{fig:J-range-influence}(a) and (b) present the
dependence of the reconstruction on the interval. In (a), when the span is
small, $J_{i, i + 1} \in [0.9, 1.1]$, the $J_{i, i + 1}$ can be correctly
estimated up to $i = 28$. While when the span is large, $J_{i, i + 1} \in
[0.8, 1.2]$, the applicability deteriorates---the estimated values $J'_{i, i +
1}$ starts deviating from $J_{i, i + 1}$ around site 16, as can be
straightforwardly read from the error defined by $\delta J_{i, i + 1} = J'_{i,
i + 1} - J_{i, i + 1}$ as shown in Fig. \ref{fig:J-range-influence}(c).

As discussed above, the reconstruction is rather robust for a long chain with
identical $J_{i, i + 1}$. When $J_{i, i + 1}$ are randomly distributed,
however, the effective distance of reconstruction is significantly shortened,
showing an anticorrelation between the distance and the distribution interval
of $J_{i, i + 1}$. Examining the spectral distribution, the anticorrelation
can also be explained by the truncation of $C_{1, n}$ as discussed for the
homogeneous chain. The distribution interval of $J_{i, i + 1}$ influences the
spectral pattern and the probability to find $C_{1, n}$ around zero. With the
distribution of random $J_{i, i + 1}$ broadened, the spectral distribution
Figs. \ref{fig:J-range-influence}(d) and (e) is no more as regular as that in
the homogeneous chain. The $C_{1, n}$ scatter to a larger range with the
increasing distribution interval of $J_{i, i + 1}$, and more $C_{1, n}$
approach zero. As discussed for the homogeneous chain, these near-zero $C_{1,
n}$ are likely to be truncated, and the resultant lost spectral components
worsen the applicability of the algorithm. In addition, it is found that the
decreasing \ $J_{i, i + 1}$ also lowers the value of $C_{1, n}$ and impedes
the parameter estimation, as can be intuitively understood since any broken
bridge hinders the probe of further sites. The above discussion also applies
when disorders of $\varepsilon_i$ are involved as the spectral distribution
also presents the similar pattern and suggests the involvement of the
localization.

\begin{figure}[h]
  \includegraphics[scale=1]{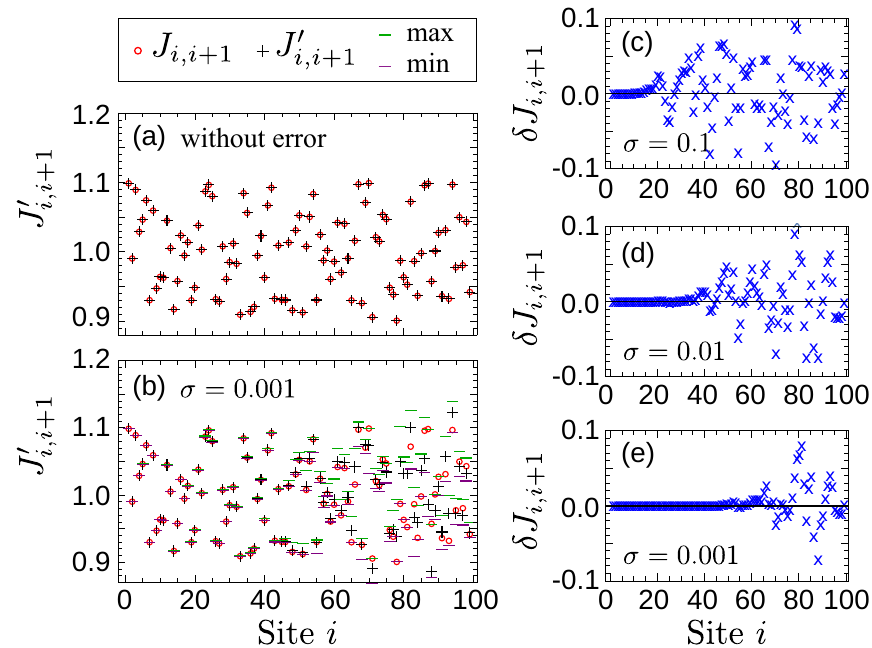}
  \caption{The estimation of coupling strengths when $\Delta \lambda_n$ is
  considered for each measurement. In a chain of 100 sites with $J_{i, i + 1}
  \in [0.9, 1.1]$, the reconstructed $J_{i, i + 1}'$ are compared with $J_{i,
  i + 1}$ in (a) when $\Delta \lambda_n = 0$. In (b), the reconstructed $J_{i,
  i + 1}'$ are shown when $\sigma = 0.001$. The errors $\delta J_{i, i + 1}$
  are presented in (c), (d) and (e) for $\sigma = 10^{- 1}$, $10^{- 2}$ and
  $10^{- 3}$, respectively.\label{fig:ensemble-influence}}
\end{figure}

Besides the influence from errors of $C_{1, n}$, the measurement of
$\lambda_n$ also affects the $J_{i, i + 1}$ reconstruction. Assuming the
actual eigenvalue is $\lambda_n$, the restricted resolution or disturbance
during the experiment may deviate the measured value from $\lambda_n$,
$\lambda_n' = \lambda_n + \Delta \lambda_n$. The fluctuation $\Delta
\lambda_n$ in each measurement results in the difference of estimated $J_{i, i
+ 1}'$, as illustrated in Fig. \ref{fig:demo-reconstruction}(d). The eventual
$J_{i, i + 1}'$ should be the average of reconstructed results after multiple
measurements. We sample 2000 random $\Delta \lambda_n$ of normal distribution,
$\Delta \lambda_n \sim \mathcal{N} (0, \sigma^2)$, to estimated the randomly
distributed \ $J_{i, i + 1} \in [0.9, 1.1]$ in a chain of $N = 100$ as shown
in Fig. \ref{fig:ensemble-influence}(a). Here, the influence by the truncation
of small-valued $C_{1, n}$ is neglected.

From the errors $\delta J_{i, i + 1}$ as shown in Fig.
\ref{fig:ensemble-influence}(c), (d) and (e), the $J_{i, i + 1}$ can be
roughly estimated up to $i = 20$, $40$ and $60$, respectively, for $\sigma =
10^{- 1}$, $10^{- 2}$ and $10^{- 3}$. While without $\Delta \lambda_n$
fluctuation [Fig. 5(a)], all $J_{i, i + 1}$ are correctly reconstructed.
Therefore, the precise measurement $\lambda_n$ is shown to be critical to the
precise reconstruction. For a longer chain, we did not find significant
deviations. However, the denser spectral distribution with more states
involved will hinder the precise retrieval of $\lambda_n$, if the instrumental
resolution of $\lambda_n$-acquisition is finite.

\section{Conclusion}\label{sec5}

The reconstruction of parameters within a partially accessible system is an
important problem when indirect probing is the only option to acquire the
desired information. In this work, the algorithm to estimate parameters in a
linear chain is investigated. Starting with the generic Schr{\"o}dinger
equation, it is confirmed that the coupling strengths can be efficiently
deduced from the recursive relation using the spectral information of only the
end site. We focus on the applicability of the algorithm, which is shown to be
highly relevant to the spectral distribution on the accessible state. Given
the errors induced by the finite signal-noise ratio, the increasing number of
truncated spectral components are shown to gradually deteriorate the
reconstruction performance. It is found that reducing the loss of spectral
components is critical to the success of the method. Even so, the partially
successful estimation in the presence of truncations shows the robustness of
the algorithm and the estimation can be conducted in a controllable way. Since
the spectral distribution is system dependent, the applicability of the method
also varies with systems. Accordingly, it is advisable to understand the
nature of the system before applying the method.

\begin{acknowledgments}
This work is supported by Shanghai Sailing Program
(16YF1412600); the National Natural Science Foundation of China (Grants No.
11420101003, No. 11604347, No. 11827806, No. 11874368 and No. 91636105).
\end{acknowledgments}

\bibliographystyle{apsrev4-1}
\bibliography{main}

\end{document}